# Reduction of periodic one-dimensional hopping model

Yunxin  Zhang*

### Abstract

In this research, methods of reducing a general periodic one-dimensional hopping model to a one- or two-state model, which keeps the basic properties of the original process, are discussed. This reduction also implies that, to some extent, many processes can be well described by simple two-state or even one-state models.

## I.   INTRODUCTION

Many biophysical [1, 2] and biochemical phenomena, especially for the motion of motor proteins [3–6], can be well described by periodic one-dimensional hopping models. In these models, the particle jumps along a periodical linear track (e.g., microtubule or filament for motor proteins kinesin, dynein and myosion [7–9]) from one binding site to next one through the sequence of $N$ mechanochemical states [10, 11]. The particle in state $j$ can jump forward to state $j + 1$ with rate $u_j$, or jump backward to state $j - 1$ with rate $w_j$. After moving $N$ sites forward or backward, the particle comes to the same mechanochemical state but shifted by a step size distance $L$ (for example, $L = 8$ nm for conventional kinesin[12] and cytoplasmic dynein [13], and $L = 36$ nm for mysion V [14]).

In one-dimensional hopping model, the motion of the particle can be described by the

_______
*Shanghai Key Laboratory for Contemporary Applied Mathematics, Centre for Computational System Biology, and School of Mathematical Sciences, Fudan University, Shanghai 200433, China.

Email: xyz@fudan.edu.cn. Phone: 86-21-55665008, Fax: 86-21-65642341

standard rate equations of occupation probabilities $p_j(t)$ [10, 15, 16]

$$\frac{\partial p_j(t)}{\partial t} = u_{j-1}p_{j-1}(t) + w_{j+1}p_{j+1}(t) - [u_j + w_j]p_j(t),$$
$$0 \leq j \leq N-1, \tag{1}$$

where $p_j(t)$ is the probability of finding the particle in state $j$ and at time $t$, which satisfies

$$p_{lN+j}(t) = p_j(t), \quad u_{lN+j} = u_j, \quad w_{lN+j} = w_j, \tag{2}$$

in which $l$ is an integer number. At steady state,

$$u_{j-1}p_{j-1} + w_{j+1}p_{j+1} - [u_j + w_j]p_j = 0, \quad 0 \leq j \leq N-1. \tag{3}$$

It's solution is

$$p_j = \frac{r_j}{R_N}, \quad 0 \leq j \leq N-1, \tag{4}$$

where

$$r_j = \frac{1}{u_j}\left[1 + \sum_{k=1}^{N-1}\prod_{i=j+1}^{j+k}\frac{w_i}{u_i}\right], \qquad R_N = \sum_{j=0}^{N-1} r_j. \tag{5}$$

This model has been extensively studied [17, 18] and the explicit formulations of mean velocity $V_N$ and (effective) diffusion constant $D_N$ had been obtained by B. Derrida [16]:

$$V_N = \frac{L\left[1 - \prod_{j=0}^{N-1}\frac{w_i}{u_i}\right]}{R_N}, \qquad D_N = \frac{L}{N}\left[\frac{LG_N + VS_N}{R_N^2} - \frac{(N+2)V}{2}\right], \tag{6}$$

where

$$S_N = \sum_{j=0}^{N-1} s_j \sum_{k=0}^{N-1}(k+1)r_{k+j+1}, \quad G_N = \sum_{j=0}^{N-1} u_j r_j s_j, \quad s_j = \frac{1}{u_j}\left[1 + \sum_{k=1}^{N-1}\prod_{i=j-1}^{j-k}\frac{w_{i+1}}{u_i}\right].$$

At the same time, in the literature of first passage time problems, the particle is assumed to jump inside a finite interval $[-(M+1), K+1]$ with absorbing boundaries. If the forward and backward jump rates in state $-M \leq n \leq K$ are $U_n, W_n$, then, the mean first passage time $\overline{T}_n$ for a particle starting at state $n$ to reach boundaries $K+1$ or $-(M+1)$, satisfies the following equations

$$\overline{T}_n = \frac{1}{U_n + W_n} + \frac{U_n}{U_n + W_n}\overline{T}_{n+1} + \frac{W_n}{U_n + W_n}\overline{T}_{n-1}, \quad \text{for } -M \leq n \leq K, \tag{7}$$

or equivalently,

$$U_n(\overline{T}_{n+1} - \overline{T}_n) + W_n(\overline{T}_{n-1} - \overline{T}_n) = -1, \quad \text{for } -M \leq n \leq K. \tag{8}$$

Combining with the absorbing boundary conditions, $\overline{T}_{-(M+1)} = \overline{T}_{(K+1)} = 0$, the explicit solutions of (8) can be obtained [19]

$$\overline{T}_n = \frac{1 + \sum_{k=-M}^{n-1} \prod_{j=-M}^{k} \frac{W_j}{U_j}}{1 + \sum_{k=-M}^{L} \prod_{j=-M}^{k} \frac{W_j}{U_j}} \left[ \sum_{k=-M}^{L} \frac{1}{U_k} + \sum_{k=-M+1}^{L} \left( \frac{1}{U_k} \sum_{i=-M}^{k-1} \prod_{j=i}^{k-1} \frac{W_{j+1}}{U_j} \right) \right]$$
$$- \left[ \sum_{k=-M}^{n-1} \frac{1}{U_k} + \sum_{k=-M+1}^{n-1} \left( \frac{1}{U_k} \sum_{i=-M}^{k-1} \prod_{j=i}^{k-1} \frac{W_{j+1}}{U_j} \right) \right] \quad \text{for } -M \leq n \leq L. \tag{9}$$

Therefore, as a special case, the mean first passage time of a particle in one-dimensional hopping model of period $N$, to complete one forward or backward mechanochemical cycle is

$$T_N = \frac{R_N}{\left[ 1 + \prod_{j=0}^{N-1} \frac{w_i}{u_i} \right]}. \tag{10}$$

Here, $T_N$ is equal to $\overline{T}_0$ in (9), but replacing $U_n, W_n$ with $u_n, w_n$, and let $M = K = N - 1$, $u_{-n} = u_{N-n}, w_{-n} = w_{N-n}$, for $0 \leq n \leq N - 1$.

In the next section, based on the explicit formulations of the mean velocity, mean first passage time and effective diffusion constant as given in (6) (10), we will discuss how to approximate a $N$-state model by a simple one-state model, and then in Section **III**, the approximation by a two-state model is addressed. The reduction of models with infinite internal states cases is discussed in Section **IV**, and finally concluding remarks are provided in Section **V**.

## II.  REDUCED ONE-STATE MODEL

If we are not interested in the intermediate states and details of the one-dimensional hopping model, and only want to know the basic biophysical and biochemical properties of the corresponding processes, such as the mean velocity, mean first passage time, or effective diffusion constant, a $N$-state model can be approximated by a simple one-state model with reduced transition rates $u_r, w_r$. In fact, in many experiments only these basic properties can be measured directly.

To a one-state model with forward and backward transition rates $u_r, w_r$, its mean velocity is $(u_r - w_r)L$, and its mean first passage time is $1/(u_r + w_r)$: see formulations in (6) and (10). So, under the assumption that the reduced one-state model has the same mean velocity and mean first passage time as those of the $N$-state model, the reduced rates $u_r, w_r$ satisfy

$$u_r - w_r = \left[ 1 - \prod_{j=0}^{N-1} \frac{w_i}{u_i} \right] \Big/ R_N, \qquad u_r + w_r = \left[ 1 + \prod_{j=0}^{N-1} \frac{w_i}{u_i} \right] \Big/ R_N, \qquad (11)$$

which implies

$$u_r = 1/R_N, \qquad w_r = \prod_{j=0}^{N-1} \frac{w_i}{u_i} \Big/ R_N. \qquad (12)$$

In fact, the same formulations have been used by some authors to simplify the complex models [20, 21], though no detailed discussion has been given. The application of such simplification indicates this method is valuable, and sometimes essential, to get meaningful explicit results.

However, it should be careful to use such reduction method. Although the reduced one-state model has the same mean velocity and mean first passage time as the original one's, it cannot be used to get the properties related to high order momentums, such as the effective diffusion constant $D_N$ and randomness $2D/VL$. In fact, the effective diffusion constant of the reduced one-state hopping model is

$$D_r = \frac{u_r + w_r}{2} L^2 = L^2 \left[ 1 + \prod_{j=0}^{N-1} \frac{w_i}{u_i} \right] \Big/ (2R_N), \qquad (13)$$

which can be obtained by the formulation in (6): with $N = 1$ and $u_0 = u_r, w_0 = w_r$. Obviously $D_r$ is different from $D_N$.

If we are more interested in the mean velocity and effective diffusion constant than the mean first passage time, we may approximate a $N$-state model by a simple one-state model, which keeps the mean velocity and effective diffusion constant. From formulations in (6) and (13), the corresponding reduced rates $u_r, w_r$ can be obtained by the following equations

$$u_r - w_r = [1 - \Gamma] / R_N,$$
$$u_r + w_r = \frac{2}{N} \left[ \frac{(1 - \Gamma) S_N}{R_N^3} + \frac{U_N}{R_N^2} - \frac{(N+2)(1-\Gamma)}{2R_N} \right], \qquad (14)$$

in which $\Gamma = \prod_{j=0}^{N-1} \frac{w_i}{u_i}$. So we have

$$
\begin{aligned}
u_r &= \frac{1}{N}\left[\frac{(1-\Gamma)\,S_N}{R_N^3} + \frac{U_N}{R_N^2} - \frac{(1-\Gamma)}{R_N}\right], \\
w_r &= \frac{1}{N}\left[\frac{(1-\Gamma)\,S_N}{R_N^3} + \frac{U_N}{R_N^2} - \frac{(N+1)(1-\Gamma)}{R_N}\right].
\end{aligned}
\tag{15}
$$

## III. REDUCED TWO-STATE MODEL

In most cases, the reduced one-state model is enough to get the basic properties of the biophysical and biochemical problems. However, if we want a simple model that can recover the mean velocity, mean first passage time, and effective diffusion constant of the original precesses at the same time, at least a two-state model should be employed.

For convenience, let $u_{ri}, w_{ri}(i=0,1)$ be the forward and backward transition rates in state $i$ of the reduced two-state model, and $u := u_{r0}u_{r1}, w := w_{r0}w_{r1}, \Sigma := u_{r0} + u_{r1} + w_{r0} + w_{r1}$. From the formulations in (6) and (10), or the results in [10, 22, 23], one can verify that the mean velocity, mean first passage time and effective diffusion constant of the two-state model are the following

$$
\begin{aligned}
V &= (u-w)L/\Sigma, \qquad T = \Sigma/(u+w), \\
D &= \frac{1}{2}\left[\frac{u+w}{\Sigma} - 2\left(\frac{u-w}{\Sigma}\right)^2 \frac{1}{\Sigma}\right]L^2 = \frac{L^2}{2T} - \frac{V^2}{\Sigma}.
\end{aligned}
\tag{16}
$$

So

$$
\begin{aligned}
u &= \frac{L+VT}{2TL}\Sigma = \frac{(L+VT)V^2}{(L^2-2DT)L}, \\
w &= \frac{L-VT}{2TL}\Sigma = \frac{(L-VT)V^2}{(L^2-2DT)L}, \\
\Sigma &= \frac{2TV^2}{L^2-2DT}.
\end{aligned}
\tag{17}
$$

If the reduced two-state model has the same mean velocity, mean first passage time and effective diffusion constant as those of the original $N$-state model, then, by formulations in

(6) and (10) , one can easily show

$$
\begin{aligned}
L^2 - 2DT &= \frac{2L^2}{N}\left(\frac{N+1-\Gamma}{1+\Gamma} - \frac{(1-\Gamma)S_N}{(1+\Gamma)R_N^2} - \frac{U_N}{(1+\Gamma)R_N}\right), \\
2TV^2 &= \frac{2(1-\Gamma)^2 L^2}{(1+\Gamma)R_N}, \\
L + VT &= \frac{2L}{1+\Gamma} \qquad L - VT = \frac{2\Gamma L}{1+\Gamma}.
\end{aligned}
\tag{18}
$$

Therefore, Eq. (17) gives

$$
\begin{aligned}
u &= \frac{N(1-\Gamma)^2}{(N+1-\Gamma)R_N^2 - U_N R_N - (1-\Gamma)S_N}, \\
w &= \frac{N(1-\Gamma)^2\Gamma}{(N+1-\Gamma)R_N^2 - U_N R_N - (1-\Gamma)S_N}, \\
\Sigma &= \frac{N(1-\Gamma)^2 R_N}{(N+1-\Gamma)R_N^2 - U_N R_N - (1-\Gamma)S_N}.
\end{aligned}
\tag{19}
$$

Finally, from $u = u_{r0}u_{r1}, w = w_{r0}w_{r1}, \Sigma = u_{r0} + u_{r1} + w_{r0} + w_{r1}$, the reduced rate constants $u_{r0}, u_{r1}, w_{r0}, w_{r1}$ can be obtained.

Obviously, there usually exist more than one reduced two-state models which has the same mean velocity, mean first passage time and effective diffusion constant as those of the original $N$-state model. To obtain a unique reduced two-state model, it should be required to keep more properties of the original $N$-state model, such as the high order momentum of displacement or first passage time, which depends on the problems that we are interested in.

## IV.   REDUCTION OF CONTINUOUS MODELS

The continuous models can be regarded as the limit of the discrete models discussed above, in which the number $N$ of internal states tends to infinity. To the continuous models, the probability density $\rho(x,t)$ of finding the particle in position (or state) $x$ and at time $t$, is governed by the following Fokker-Planck equation [11, 15, 24]

$$
\frac{\partial \rho}{\partial t} = \frac{\partial}{\partial x}\left(\frac{\rho}{\xi}\frac{\partial \Phi}{\partial x} + D\frac{\partial \rho}{\partial x}\right), \qquad -\infty \leq x \leq +\infty,
\tag{20}
$$

in which $\xi$ is the drag coefficient, and $D$ is the free diffusion constant which satisfies the Einstein relation $D\xi = k_B T$, here $k_B$ is Boltzmann constant and $T$ is absolute temperature.

$\Phi$ is a tilted periodic potential, $\Phi(x - L) = \Phi(x) + FL$. At steady state, the mean velocity can be obtained by the following formulation

$$
\begin{aligned}
V &= \frac{d}{dt} \int_{-\infty}^{\infty} x \rho(x, t) dx = \int_{-\infty}^{\infty} x \frac{\partial}{\partial x} \left( \frac{\rho}{\xi} \frac{\partial \Phi}{\partial x} + D \frac{\partial \rho}{\partial x} \right) dx \\
&= -\int_{-\infty}^{\infty} \left( \frac{\rho}{\xi} \frac{\partial \Phi}{\partial x} + D \frac{\partial \rho}{\partial x} \right) dx.
\end{aligned}
\tag{21}
$$

Under the similar assumption that is used in the derivation of formulations in (6), the explicit expression of $V$ can be obtained [15]

$$
V = \frac{D(1 - e^{-\beta FL})L}{\int_0^L e^{-\beta \Phi(x)} \left( \int_x^{x+L} e^{\beta \Phi(y)} dy \right) dx} = \frac{D(1 - e^{-\beta FL})L}{\int_0^L e^{\beta \Phi(x)} \left( \int_{x-L}^x e^{-\beta \Phi(y)} dy \right) dx},
\tag{22}
$$

in which $\beta = 1/k_B T$.

At the same time, the mean first passage time $T(x)$ of particle starting at location $x$ to reach boundaries $x = L$ or $x = -L$ is governed by the following differential equation [25]

$$
D \frac{\partial^2 T(x)}{\partial x^2} - \frac{1}{\xi} \frac{\partial \Phi(x)}{\partial x} \frac{\partial T(x)}{\partial x} = -1.
\tag{23}
$$

Combining with the absorbing boundary conditions, i.e. $T(L) = T(-L) = 0$, the mean first passage time $T(0)$ for the particle to complete one forward or backward step, can be obtained by the following formulation

$$
\begin{aligned}
T(0) &= \frac{\left( \int_{-L}^L e^{\beta \Phi(y)} dy \right) \left[ \int_0^L e^{\beta \Phi(y)} \left( \int_0^y e^{-\beta \Phi(z)} dz \right) dy \right] - \left( \int_0^L e^{\beta \Phi(y)} dy \right) \left[ \int_{-L}^L e^{\beta \Phi(y)} \left( \int_0^y e^{-\beta \Phi(z)} dz \right) dy \right]}{D \int_{-L}^L e^{\beta \Phi(y)} dy} \\
&= \frac{\left( \int_{-L}^0 e^{\beta \Phi(y)} dy \right) \left[ \int_0^L e^{\beta \Phi(y)} \left( \int_0^y e^{-\beta \Phi(z)} dz \right) dy \right] - \left( \int_0^L e^{\beta \Phi(y)} dy \right) \left[ \int_{-L}^0 e^{\beta \Phi(y)} \left( \int_0^y e^{-\beta \Phi(z)} dz \right) dy \right]}{D \int_{-L}^L e^{\beta \Phi(y)} dy}.
\end{aligned}
\tag{24}
$$

$$
-\int_{-L}^0 e^{\beta \Phi(y)} \left( \int_0^y e^{-\beta \Phi(z)} dz \right) dy = \int_0^L e^{\beta \Phi(y)} \left( \int_y^L e^{-\beta \Phi(z)} dz \right) dy,
$$

$$
\int_{-L}^0 e^{\beta \Phi(y)} dy = e^{\beta FL} \int_0^L e^{\beta \Phi(y)} dy \qquad \int_{-L}^L e^{\beta \Phi(y)} dy = (1 + e^{\beta FL}) \int_0^L e^{\beta \Phi(y)} dy.
$$

So

$$
\begin{aligned}
T(0) =& \frac{e^{\beta FL}\left[\int_0^L e^{\beta\Phi(y)}\left(\int_0^y e^{-\beta\Phi(z)}dz\right)dy\right]+\left[\int_0^L e^{\beta\Phi(y)}\left(\int_y^L e^{-\beta\Phi(z)}dz\right)dy\right]}{D(1+e^{\beta FL})}\\
=& \frac{\left[\int_0^L e^{\beta\Phi(y)}\left(\int_0^y e^{-\beta\Phi(z)}dz\right)dy\right]+e^{-\beta FL}\left[\int_0^L e^{\beta\Phi(y)}\left(\int_y^L e^{-\beta\Phi(z)}dz\right)dy\right]}{D(1+e^{-\beta FL})}\\
=& \frac{\left[\int_0^L e^{\beta\Phi(y)}\left(\int_0^y e^{-\beta\Phi(z)}dz\right)dy\right]+\left[\int_0^L e^{\beta\Phi(y)}\left(\int_{y-L}^0 e^{-\beta\Phi(z)}dz\right)dy\right]}{D(1+e^{-\beta FL})}\\
=& \frac{\int_0^L e^{\beta\Phi(y)}\left(\int_{y-L}^y e^{-\beta\Phi(z)}dz\right)dy}{D(1+e^{-\beta FL})}.
\end{aligned}
\tag{25}
$$

Therefore, if the reduced one-state model, with forward and backward transition rates $u_r, w_r$, has the same mean velocity and mean first passage time as those of the continuous model, then

$$
u_r - w_r = \frac{V}{L}, \qquad u_r + w_r = \frac{1}{T(0)}.
$$

So, the reduced rates $u_r, w_r$ can be obtained as follows

$$
u_r = \frac{D}{\int_0^L e^{\beta\Phi(y)}\left(\int_{y-L}^y e^{-\beta\Phi(z)}dz\right)dy}, \qquad w_r = \frac{De^{-\beta FL}}{\int_0^L e^{\beta\Phi(y)}\left(\int_{y-L}^y e^{-\beta\Phi(z)}dz\right)dy}.
\tag{26}
$$

By formulations in (12) and (26), one can find some relations between the discrete models (1) (8) and the continuous models (20) (23), and some relations between the transition rates $u_i, w_i$ and the potential $\Phi(x)$, for more details see [26].

Using the same method which has been employed in the the discussion of the discrete cases, we also can get a reduced one-state model which has the same mean velocity and effective diffusive constant $D_{eff}$ as those of the continuous model, or get a reduced two-state model which preserves all the three physical quantities. Here, the effective diffusion constant $D_{eff}$ is defined as follows

$$
D_{eff} := \frac{1}{2}\lim_{t\to\infty}\left[\frac{d\overline{x^2}}{dt} - \frac{d(\bar{x})^2}{dt}\right],
\tag{27}
$$

with

$$
\overline{x^k}(t) := \int_{-\infty}^\infty x^k \rho(x,t)d\,x \qquad k = 1,2.
$$

The explicit formulation of $D_{eff}$ can be found in [15] or we can just use the limit of $D_N$ in (6) (see [26]).

## V.   CONCLUDING REMARKS

In this paper, the methods of how to approximate a $N$-state model by a simple one-state or two-state model, which keeps some of the basic properties of the original problems are presented. These methods can be used to obtain explicit results for some complex biophysical and biochemical precesses [20, 21]. The discussion in this paper also indicates that, in many cases, a simple two-state model or even one-state model is enough to describe a complex process if we are only interested in its basic properties, such as the mean velocity, mean first passage time and effective diffusion constant. Actually, in many cases, such basic quantities are the only ones that can be measured directly in the experiments. Recent discussion about a modified one-dimensional hopping model indicates that, the method in the paper also can be used to more general hopping models, in which the particle in state $i$ has more than two choices (states $i-1$ and $i+1$) to jump out its present state [27].

### Acknowledgments

This work is funded by the National Natural Science Foundation of China (Grant No. 10701029).

----

[1] S. Alexander, J. Bernasconi, W. R. Schneider, R. Biller, W. G. Clark, G. Grüner, R. Orbach, and Zettl. Frequency-dependent charge transport in a one-dimensional disordered metal. *Phys. Rev. B*, 24:7474–7477, 1981.

[2] S. Alexander, J. Bernasconi, W. R. Schneider, and R. Orbach. Excitation dynamics in random one-dimensional systems. *Rev. Mod. Phys.*, 53:175–198, 1981.

[3] N. J. Carter and R. A. Cross. Mechanics of the kinesin step. *Nature*, 435:308–312, 2005.

[4] Anatoly B. Kolomeisky and Michael E. Fisher. Periodic sequential kinetic models with jumping, branching and deaths. *Phys. A*, 279:1–20, 2000.

[5] G. Lattanzi and A. Maritan. Master equation approach to molecular motors. *Phys. Rev. E*, 64:061905, 2001.

[6] Y. Zhang. Three phase model of the processive motor protein kinesin. *Biophys. Chem.*, 136:19–22, 2008.

[7] Koen Visscher, Mark J. Schnitzer, and Steven M. Block. Single kinesin molecules studied with amolecular force clamp. *Nature*, 400:184–189, 1999.

[8] S. Toba, T. M. Watanabe, L. Yamaguchi-Okimoto, Y. Y. Toyoshima, and H. Higuchi. Overlapping hand-over-hand mechanism of single molecular motility of cytoplasmic dynein. *Proc. Natl. Acad. Sci. USA*, 103:5741–5745, 2006.

[9] R. D. Vale. The molecular motor toolbox for intracellular transport. *Cell*, 112:467–480, 2003.

[10] M. E. Fisher and A. B. Kolomeisky. Molecular motors and the forces they exert. *Physica A*, 274:241–266, 1999.

[11] Y. Zhang. The efficiency of molecular motors. *J. Stat. Phys.*, 134:669–679, 2009.

[12] C. M. Coppin, J. T. Finer, J. A. Spudich, and R. D. Vale. Detection of sub-8-nm movements of kinesin by high-resolution optical-trap microscopy. *Proc. Natl. Acad. Sci. USA*, 93:1913–1917, 1996.

[13] Arne Gennerich, Andrew P. Carter, Samara L. Reck-Peterson, and Ronald D. Vale. Force-induced bidirectional stepping of cytoplasmic dynein. *Cell*, 131:952–965, 2007.

[14] G. Cappello, P. Pierobon, C. Symonds, L. Busoni, J. C. M. Gebhardt, M. Rief, and J. Prost. Myosin V stepping mechanism. *Proc. Natl. Acad. Sci. USA*, 104:15328–15333, 2007.

[15] Y. Zhang. Derivation of diffusion coefficient of a brownian particle in tilted periodic potential from the coordinate moments. *Phys. Lett. A*, 373:2629–2633, 2009.

[16] B. Derrida. Velocity and diffusion constant of a periodic one-dimensional hopping model. *J. Stat. Phys.*, 31:433–450, 1983.

[17] I. Webman. Effective-medium approximation for diffusion on a random lattice. *Phys. Rev. Lett.*, 47:1496–1499, 1981.

[18] J. Machta. Generalized diffusion coefficient in one-dimensional random walks with static disorder. *Phys. Rev. B*, 24:5260–5269, 1982.

[19] P. A Pury and M. O Cáceres. Mean first-passage and residence times of random walks on asymmetric disordered chains. *J. Phys. A: Math. Gen.*, 36:2695–2706, 2003.

[20] A. B. Kolomeisky, E. B. Stukalin, and A. A. Popov. Understanding mechanochemical coupling in kinesins using first-passage-time processes. *Phys. Rev. E*, 71:031902, 2005.

[21] Denis Tsygankova and Michael E. Fisher. Kinetic models for mechanoenzymes: Structural aspects under large loads. *J. Chem. Phys.*, 128:015102, 2008.

[22] M. E. Fisher and A. B. Kolomeisky. Simple mechanochemistry describes the dynamics of

kinesin molecules. *Proc. Natl. Acad. Sci. USA*, 98:7748–7753, 2001.

[23] A. B. Kolomeisky and M. E. Fisher. A simple kinetic model describes the processivity of myosin-v. *Biophys. J.*, 84:1642–1650, 2003.

[24] J. Howard. *Mechanics of Motor Proteins and the Cytoskeleton.* Sinauer Associates and Sunderland, MA, 2001.

[25] H. M. Taylor and S. Karlin. *An Introduction to Stochastic Modeling.* Academic Press, San Diego, 1998.

[26] Y. Zhang. Limit properties of periodic one dimensional hopping model. *Chinese J. Chem. Phys.*, 23:65–68, 2010.

[27] Y. Zhang. Properties of modified periodic one-dimensional hopping model. *in preparation*, 2010.